\documentclass[preprint2]{aastex}
\usepackage{natbib}
\newcommand{\kms}{\hbox{km\,s$^{-1}$}}

\shorttitle{Balmer line variability in $\eta$~Car}
\shortauthors{Weis et al.}

\begin{document}
\title{VLT-UVES observations of the Balmer line
variations of $\eta$~Carinae during the 2003
spectroscopic event}

\author{K.\ Weis\altaffilmark{1}}
\affil{Astronomisches Institut, Ruhr-Universit\"at Bochum,
Universit\"atsstr. 150, 44780 Bochum,  Germany}
\altaffiltext{1}{Lise-Meitner fellow}
\email{kweis@astro.rub.de}

\author{O.\ Stahl}
\affil{Landessternwarte Heidelberg, K\"onigstuhl, 69117  Heidelberg,
Germany}
\email{O.Stahl@lsw.uni-heidelberg.de}

\author{D.\ J.\ Bomans}
\affil{Astronomisches Institut, Ruhr-Universit\"at Bochum,
Universit\"atsstr. 150,
44780 Bochum,  Germany}
\email{bomans@astro.rub.de}

\author{K.\ Davidson}
\affil{Astronomy Department, University of Minnesota, Minneapolis, MN 55455}
\email{kd@astro.umn.edu}

\author{T.\ R.\ Gull}
\affil{Code 681,NASA/GSFC, Greenbelt, MD 20771}
\email{Theodore.R.Gull@nasa.gov}

\and

\author{R.\ M.\ Humphreys}
\affil{Astronomy Department, University of Minnesota, Minneapolis, MN 55455}
\email{roberta@aps.umn.edu}

\begin{abstract}

We present high spectral resolution echelle observations of the
Balmer line variations during the 2003.5 ``spectroscopic  event'' of
$\eta$~Carinae. Spectra have been recorded of both  $\eta$~Carinae
and the Homunculus at the FOS4 position in its SE lobe. This spot
shows a reflected stellar spectrum which is less contaminated by
nebular emission lines than ground-based observations of the central
object itself. Our observations show that the spectroscopic event is
much less pronounced at this position than when seen directly on
$\eta$~Car using HST/STIS. Assuming that the reflected spectrum is
indeed latitude dependent this indicates that the spectral changes
during the event seen pole-on (FOS4) are different from those closer
to the equator (directly on the star). In contrast to the spectrum
of the star, the scattered spectrum of FOS4 always shows pronounced
P~Cygni absorption with little variation across the ``spectroscopic
event''. After that event an additional high-velocity absorption
component appears. The emission profile is more peaked at FOS4 and
consists of at least 3 distinct components, of which the reddest one
shows the strongest changes through the event. The data seem to be
compatible with changes in latitudinal wind structure 
of a single star, with or without the help of a
secondary star,  or the  onset of a shell ejection during
the spectroscopic event.
\end{abstract}


\section{Introduction}

$\eta$~Car is among the most massive and luminous unstable stars of our
Galaxy. During its ``great eruption'' (about 1838-1858) it became
one of  the visually brightest stars in the sky.
Today  the star itself is hidden
in a bipolar nebula known as the {\it Homunculus\/}.
It  is, due to the dust within the expanding lobes
scattering light from $\eta$~Car, mainly a reflection nebula
\citep[for example][]{1956Obs....76..154T, 1961Obs....81..102T,
1967MNRAS.135..275V, 1979MNRAS.187..761W, 1993PASAu..10..338A}.

Temporary spectral changes
were recorded on several occasions beginning in 1948
\citep[see references cited
by][]{1999ecm..conf..221D}, and occur approximately every
 5.5 years \citep{1996ApJ...460L..49D}.
The same timescale had earlier appeared in infrared photometry
\citep{1994MNRAS.270..364W}, and was confirmed by successful
predictions of spectroscopic events around 1997.9 and 2003.5.  In
each case the excitation and ionization level of the spectrum
decreased for several months \citep[see,
e.g.][]{1984A&A...137...79Z, 1999ecm..conf..304D,
1999ecm..conf..227D, 1999ecm..conf..236M,1999ecm..conf..221D}.  At
the same time the X-ray emission first peaks  and then plummets
during the event \citep{1999ecm..conf..266I}.

The 5.54-year periodicity led many authors to suggest that
$\eta$~Car has a companion star \citep{1997NewA....2..107D,
1997NewA....2..387D, 1999ApJ...524..983I}.
\citet{1984A&A...137...79Z} have proposed earlier that each event
involves a shell ejection or similar event. The periodicity suggests
that a companion star may regulate the cycle, but neither of these
conjectures has been confirmed.  In general the phenomenon has not
yet been explained.

$\eta$~Car was observed regularly at high spatial resolution with
HST/STIS from 1998 through 2002 to separate the star from its nearby
ejecta which contaminates all ground-based spectroscopy. The HST
Treasury Project carried out an intensive monitoring with more
frequent observations during its 2003 ``spectroscopic event''. To
complement this program with improved temporal sampling during the
``event'' we observed $\eta$ Car with the {\bf U}ltraviolet and {\bf
V}isual {\bf E}chelle {\bf S}pectrograph {\sc UVES}, at the  {\it
European Southern Observatory (ESO\/)} between December 2002 and
March 2004. We observed  the reflected spectrum of the star at the
FOS4 position \citep[about 2\farcs6 south and 2\farcs8 east of the
star;][]{1999ecm..conf..107H} which allows us to view the star and
the ``event'' from a different angle. At FOS4 there is also only a
modest amount of contamination by nebular emission from the vicinity
of the star \citep[mainly the Weigelt knots at a distance of only 0\farcs1
to 0\farcs3][]{1986A&A...163L...5W, 1988A&A...203L..21H}. Spectra
taken on the star's position and at FOS4 are shown in Fig.
\ref{fig_star} for comparison. At FOS4, spectral changes in
the stellar spectrum can be observed after a light-travel delay time
of roughly 3 weeks. We also observed the star itself less frequently
but coordinated with the HST/STIS Treasury Project.

The scattering surface in the SE lobe at the FOS4 position `sees'
the star from a roughly polar direction \citep{2003ApJ...586..432S}
while  our direct line of sight sees a spectrum representative of
the intermediate latitudes \citep{2001AJ....121.1569D}. Thus these
two lines of sight allow us to follow the event from different
angles. For example, in the HST/STIS spectra of the central star, a
deep P~Cygni H$_{\alpha}$ absorption was present during the 1997-98
event but disappeared a few months later. Meanwhile, however, the
P~Cygni absorption did not disappear in the star's spectrum
reflected by dust in the southeast Homunculus lobe. A possible
explanation for the P~Cygni absorption discrepancy is that
$\eta$~Car's wind may be denser toward its poles, except during each
spectroscopic minimum---as discussed by \citet{2003ApJ...586..432S}.
In any case, some important observational questions arise. To what
extent does the reflected ``polar" H$_{\alpha}$ profile differ from
the feature that we observe directly? Does the polar H$_{\alpha}$
emission vary during a spectroscopic event, and, if so, in what
manner? Here we present partial answers to these questions, and we
describe some important details omitted by Smith et al. A
companion paper by Davidson et al. describes the  HST/STIS Balmer
line observations of the central star during 2003 which
significantly  differed from the previous one observed in 1997-98.

\section{The observations}

The data presented here are based on observations obtained  in
service mode between December 2002 and March 2004 with UVES at the
Nasmyth platform B of ESO's VLT UT2 (Kueyen) on Cerro Paranal,
Chile\footnote{Programs: 70.D-0607(A), 71.D-168(A), 72.D-052(A); PI:
Weis}. For all observations the standard settings DIC1, 346+580
(blue arm centered at 3460\,\AA, red arm centered at 5800\,\AA) and
DIC2, 437+860 (blue arm centered at 4370\,\AA\ and red arm  at
8600\,\AA) were used. The observed wavelength range extends
continuously from 3\,100\,\AA\ to 10\,200\,\AA\ except for small
gaps due to the space between the two CCDs of the detector mosaic in
the red channel. The primary  objective  was to observe the
reflected stellar spectrum at the FOS4 position (about 2\farcs6
south and 2\farcs8 east of the star) in the Homunculus. The slit was
aligned across the SE lobe of the Homunculus at a position angle of
160$\degr$. To monitor the 2003 event, spectra were taken about
every week during the event (mid May to end of July) and every month
before  and  after the event time. No spectra could  be taken
between mid August and late November due  to the very low elevation
of $\eta$~Car at Paranal during this time of the year. Integration
times were between 1\,sec (on the star) and 770\,sec on the FOS4
position.

For a reference spectrum preceeding the ``spectroscopic event'', we
used spectra obtained during the UVES commissioning (Dec. 21, 1999)
from the ESO archive. They were obtained with the identical
set-up as our spectra, but with different slit orientation. The
position angle of the slit was 45$\degr$, i.e.\ about perpendicular
to the axis of symmetry of the Homunculus with offset  to the FOS4
center. Exposure times for these spectra  were 60\,sec.
Observational dates of all spectra used for this analysis are listed
in Table~\ref{tab_obs}.

One short and one long exposure were obtained with the DIC1 (346+580
setting) to record both, the extremely bright  H$_{\alpha}$ emission
and the fainter lines with good signal-to-noise ratio. The slit
width of all observations was 0\farcs3 and 0\farcs4 in the red and
blue range, respectively, resulting in a spectral resolution of
80\,000 in the blue  and 110\,000 in  the  red arm.  The pixel
scales and slit lengths were, respectively, 0\farcs246/pixel and
7\farcs6 in the blue and 0\farcs182/pixel and 11\farcs8 in the red
arm.  The data were reduced and two-dimensional spectra extracted
using mostly the standard ESO pipeline software for UVES\@.  An
exception was the order-merging procedure where the ESO software did
not produce satisfactory results for our data, mainly because the
merging of overlapping echelle orders led  to below-optimum S/N.
 Therefore, the order merging was carried out using software
developed at the LSW Heidelberg \citep{1999oisc.conf..331S}. All
spectral frames were converted after reduction to the same (heliocentric)
wavelength scale.

Each spectrum presented here is an average of an area of about
0\farcs55 along the slit (2 rows in the blue  and 3 rows in the red
long slit spectra) extracted at the FOS4 position. We searched for
the best extraction region in the longslit, but had to settle for
this small extraction region on FOS4 to avoid smearing due to
velocity shifts across the lobe.  \citet{2003ApJ...586..432S}
encountered similar limits using HST/STIS spectra. All spectra
plotted in this paper are normalized to continuum being unity. Our
July 26 spectrum shows the minimum in the intensities of the high
excitation lines, such as He\,{\sc i} 6678\,\AA\ and therefore
represents the best ``event spectrum" in our data. The November
spectrum indicates that the affected lines  brightened significantly
afterwards.

\section{Discussion}

\subsection{The VLT UVES spectra}

The time variation of the H$_{\alpha}$ line at FOS4 is given in
Figs. \ref{fig_ha} and \ref{fig_shift}. They seem to be composites of
several components. Most obvious are at least three principal
emission components which are preliminarily fit by Gaussians plus a
P~Cygni absorption. All Gaussian profiles themselves may again be
composites of several components. It is also possible that a broader
absorption (e.g. on the  blueshifted side) causes one component to
appear as two separate ones. However, for simplicity we used  three
Gaussian. The centroids of the Gaussian are roughly at $-130$\,\kms
(FWHM $\sim 180$\,\kms), 70\,\kms\ (FWHM $\sim 230$\,\kms) and
300\,\kms (FWHM $\sim 230$\,\kms). The P~Cygni absorption is
centered at $-$340\,\kms.
For a comparison we show in Fig. \ref{fig_ha48} the time
variation of the H$_{\alpha}$ line 4\farcs8 north-west of FOS4,
closer  to the star.

During the event the H$_{\alpha}$ profile at FOS4 changes, and all
components decrease in intensity. The most notable change is in the
300\,\kms-profile---the red hump---which disappears nearly
completely. The red  hump was very prominent in the 1999 spectrum
and reappeared after the event in November 2003 and is present still
in March 2004 (see Figs.  \ref{fig_ha}, \ref{fig_shift}). Note that
the red hump was already weak in December 2002, long before the
event started, as evidenced by brightening in the X-ray
\citep{2003IAUC.8160....3C} or IR domains
\citep{2003IAUC.8160....2W, 2004MNRAS.tmp..116W}.

The second most important change in the FOS4 H$_{\alpha}$ spectra is
in the P~Cygni absorption profile. Before and during the  event
there is almost {\it no} change in the P~Cygni profile. In the more
recent VLT spectra (November 2003 -- March 2004), however a second
component at a higher  velocity---centered at about $-500$ to
$-600$\,\kms---appears. This high velocity component is first
visible in the November spectrum, is  deepest in January and weaker
again in the  March 2004 spectrum. Its appearance cannot be
explained by the light travel time, but rather shows that
this change occured later and not necessarily in causal
connection to the event. 

The travel delay time from the star to FOS4 can be estimated from
the radial velocity of the emission lines. Due to the expansion of
the Homunculus the lines are blue-shifted compared to the same lines
seen in the star's spectrum. This method however is depending on a
decent model of the expansion of the Homunculus  and a  good
measurement of the radial velocities. We used several  sharper [Fe
{\sc ii}] lines and derive a radial velocity difference of about
93\,\kms\ between the  star and FOS4. This yields a light travel
time of approximately 20 days.

The blue emission  component shows only very small changes. Of
particular interest is the central 70\,\kms-emission component. It
is intermediate in terms of variation between the red and the blue
component.

The H$_{\beta}$  and H$_{\eta}$ lines show in general the same
behavior as the H$_{\alpha}$ line (Figs. \ref{fig_hb} and
\ref{fig_hh}). The development of the red hump, however, seems to be
less strong going from H$_{\alpha}$ to H$_{\eta}$. Also the shape of
the high velocity P~Cygni component changes from H$_{\alpha}$ where
it is more straight to H$_{\eta}$ were a clear second dip is
present.


\subsection{Comparison with the HST/STIS spectra}

\begin{figure}[t]
\includegraphics[angle=270,width=8cm]{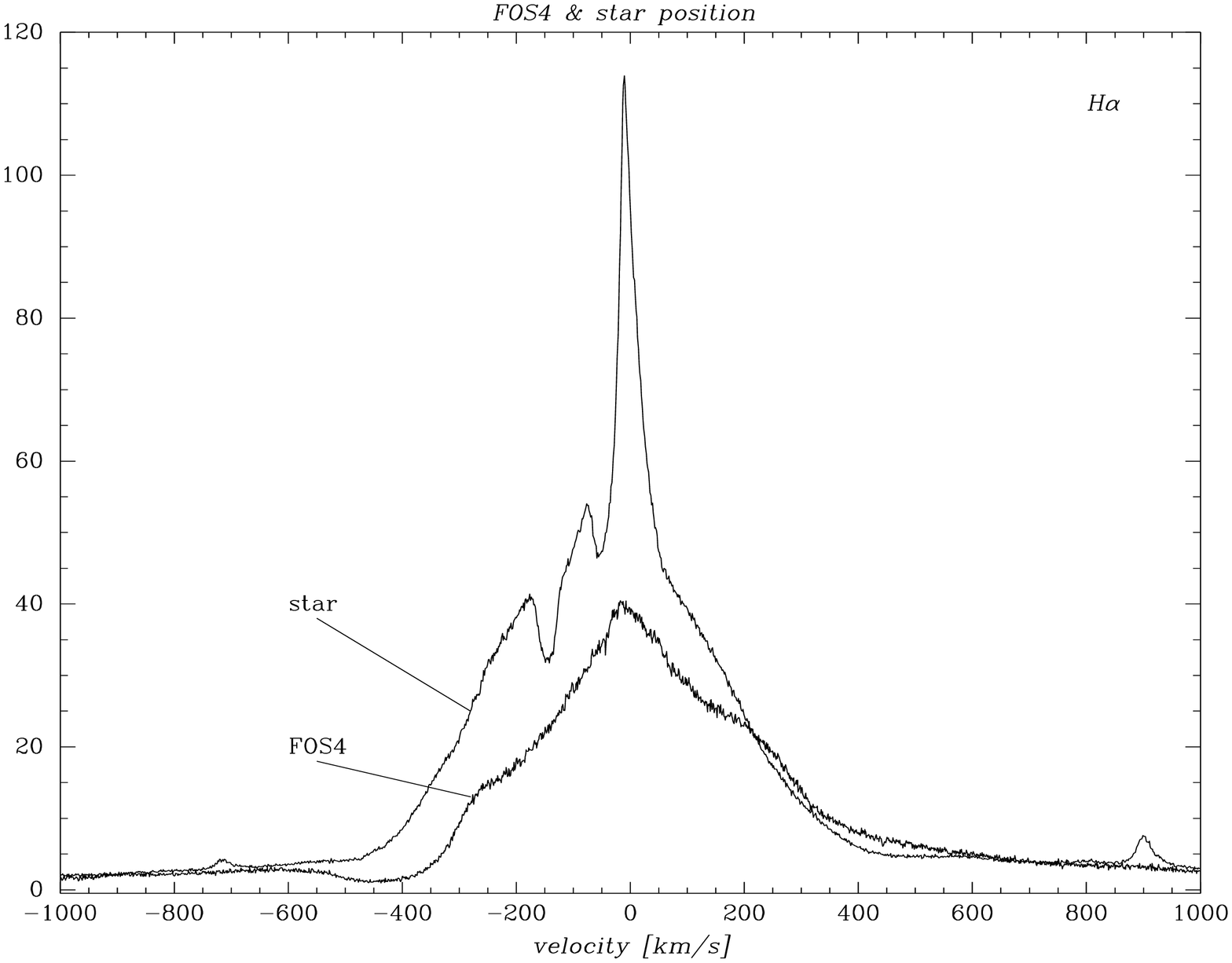}
\hspace{8cm}

\caption{Difference of ground-based spectra taken at the position of
the star and taken on the FOS4 position. The narrow and very strong
emission peaks originate from  the so called Weigelt knots which are
close ($<$0\farcs3)  to the star and are not resolved in ground
based observations. Since 1998, the star brightened considerably
while the ejecta did not, based on STIS high spatial resolution
spectra. The FOS4 spectrum plotted has been shifted by $-$93\,\kms\
to match the center of the star's spectrum. The shift is due to the
expansion of the Homunculus.} \label{fig_star}
\end{figure}

Several differences are seen in comparison with the  H$_{\alpha}$
profile profile in the HST/STIS spectra described in the
Davidson et al. 
First a sharp absorption component roughly at
$-$150\,\kms\ is seen in the HST spectra, taken directly at the
star. This component is absent in FOS4 observations, but appears  in
our spectra 
in all positions up to 1\farcs3 north west of FOS4, 
as well as in a few observations
 we made centered on 
$\eta$~Car (see Fig. \ref{fig_star}). We therefore confirm this
component to be present in an extended area (roughly 3\arcsec radius)
around the star in the  south east lobe.

Relative to the underlying continuum level, and within the
uncertainties of measurement, the UVES and HST data both show an
equally strong long-wavelength  wing at velocities above
$+600$\,\kms. This is reassuring, since wings due to Thompson
scattering should be approximately isotropic in their appearance.
However the line profile is not symmetric and the broadening is much
larger on the redshifted side. This  has been observed in supernovae
\citep{1977SvAL....3..241C}, too. A detailed analysis of the
scattering processes within the Homunculus is definitely needed to
better understand this asymmetry, but lies beyond the scope  of this
paper and will therefore be addressed  elsewhere. It is nevertheless
important since high velocity components in the P~Cygni absorption
might not be visible, or be much deeper due to the presence of
Thompson scattering wings.

The shape of the H$_{\alpha}$ profile in the HST spectra recorded
 during the 2003
event is mainly flat-topped and has changed since the last event in
1998 in which the shape was more rounded (see Davidson et al.). 
Contrary to this H$_{\alpha}$ profile we
observe at FOS4 (as discussed above) a more complex structure of the
emission. This is not an effect of the spectral resolution, but an
intrinsic difference of the two sight lines.

Before and during the  event there is little or no change in the
P~Cygni profile as viewed from the FOS4 position, the absorption
trough is rather stationary at $-$340\,\kms, and only marginally
deepens and broadens during the event (by about 50\,\kms; Figs.
\ref{fig_ha} and \ref{fig_shift}). This is in remarkable contrast to
the HST observations. There, the P~Cygni absorption is absent before
the event, develops during the event and lies at about $-$500\,\kms.
Taking the expansion velocity of the  Homunculus at the FOS4
position  into account, this is similar to the
 central velocity of the P~Cygni absorption
observed in FOS4. Here the absorption  lies at $-340$\,\kms\ but is
not corrected for the expansion. For a comparison  of our  FOS4
velocities  with measurements from the star, one has  therefore to
subtract $\sim 93$\,\kms. Consequently the P~Cygni absorption would
lie around   $-433$\,\kms.

The central velocity component (sharp peak at 70\,\kms, see  
e.g. Fig. \ref{fig_ha}) is the second major difference
between the UVES spectra at FOS4 and the STIS spectra of the star.
Without this component, the H$_{\alpha}$ profile at FOS4 resembles
the H$_{\alpha}$ profile of the star, if we ignore the much more
pronounced amplitude variation in the red hump  observed at FOS4.
This can be understood if the central component is reflected
emission from ejecta near  the  star visible at the FOS4 position.
The peaked emission might come from a gas clump, in the vicinity of
the star which would have to be bright. One other quite natural
explanation would  be that the peak emission is reflected and
scattered light of one or more of the Weigelt knots. The Weigelt
knots show extremely  bright and narrow emission lines
\citep{1995AJ....109.1784D}. Scattering at the thick walls of the
Homunculus would broaden these  lines and decrease their  intensity.
Fig. \ref{fig_star} shows that velocity of the peak emission in FOS4
agrees well with  the radial velocity of the Weigelt knots seen at
the star. Last but not least might the peak emission have its origin
in an  area called the {\it Little Homunculus\/}
\citep{2003AJ....125.3222I}. Similar in  morphology the Little
Homunculus is a small nebula ($\sim$ 2\arcsec\ in diameter) within
the Homunculus. The width of the peaked line would fit to the
expansion velocity of the Little Homunculus. 

\subsection{Global patterns}

The observations can be summarized as follows: The scattered spectrum
of the Homunculus, which is representative of the more polar region of
$\eta$~Car, is changing far less than the STIS spectrum of the core,
which is representative of  a more equatorial spectrum (more precise it
sees the star at an angle of 45$\degr$).
During the
``event'', the spectra of pole and equator become more similar.  The
change in the emission part may be partly due to changes in the
equatorial region.

It  has been shown  that the Balmer lines  seen at FOS4 show a
profile somewhat different from that seen at the  star. The
pronounced flat-topped profile of the STIS spectra is not observed
in the FOS4 UVES spectra, either because we are really seeing
different stellar profiles viewing the star more pole-on or the peak
central emission is an  additional emission superimposed on a more
flat-topped profile as seen  in the HST/STIS data. The first
scenario would imply a different appearance of the event viewed  at
the equator and viewed from a more  pole-on direction. In the second
case---the peak emission  is the result from a circumstellar
feature---the event looks  very much alike from both viewing angles.
In both directions a  flat-topped profile is seen, but in FOS4 with
the additional central emission peak. Note however that the event
view from FOS4 still is different as the P~Cygni lines are always
present as seen  from the FOS4 but change as seen from the star.

In Fig. \ref{fig_hstvlt} we have smoothed our VLT UVES FOS4 spectra
to the same spectral resolution as the HST/STIS data. This figure is
directly comparable to Fig. 1c in Davidson et al., with the same
line styles used for a dataset taken at nearly the same dates, but
taking into account the light travel time. With this delay, however,
observations should have been done monthly (as it indeed was
planned) between August and November 2003 to   compare to the STIS
observations in September 2003. Since no VLT spectra exist between
August and November 2003, we chose the last spectrum before and the
first spectrum after this gap to match Fig. 1c in Davidson et al.\
as well as possible with our data. Fig. \ref{fig_hstvlt} clearly
shows that the flat-topped profile seen in the HST spectra are not
seen in the VLT FOS4 spectra. The flat-topped profile therefore is
not an effect of the lower spectral resolution of the HST spectra,
but is indeed a  real difference between the spectrum  seen at the
star.

\section{Conclusions}

The observations  presented here show that the spectroscopic event
of $\eta$~Carinae in 2003 underwent different spectral changes when
observed from the FOS4 vantage point compared to  the line of sight
to the  star. While a flat-topped H$_{\alpha}$ profile  was found in
the HST/STIS spectra on the star position, the FOS4 spectra show a
multi-component profile  with distinct peaks. If we  assume  that
the emission observed at FOS4 is indeed only a scattered spectrum
originating from the central source and does not include emission
from other sources seen by FOS4, these observations favor a model in
which the spectroscopic event is intrinsically different at
different stellar latitudes. So the wind of $\eta$~Car appears to be
non-spherical, and the event is also non-spherical with the polar
region being affected quite differently from the equatorial region.
If the peaked emission is  of circumstellar origin the  event seen
more pole-on is only distinguished by the change of the P~Cygni
absorption profile which is present at the star but not prominent at FOS4.

The most prominent change at FOS4 is the gradual development of the
red hump, an  emission  structure. Some additional emission must be
present here, either from a companion star,  additional stellar wind
or an ejected shell. The same explanation can be put forward for the
relatively narrow central H$_{\alpha}$ component in the UVES FOS4
spectra, if it is not circumstellar. Also the appearance of an
additional  high-velocity P~Cygni component hints to new  material
between us and (the pole region of) the central source.  This
absorption component can again be the result of an increasingly
faster and denser stellar wind or a shell ejected with this
velocity.

The fact that  P~Cygni absorption is always present at FOS4 but only
appears in the line of sight to the star during the  event as seen
in the high spatial resolution HST observations has another
implication on the density structure around the central  object.
Note in this context that  P~Cygni absorption is always present in
the HST observations of the  higher Balmer lines. The explanation
for this may be  that each Balmer line is formed at a different
distance from the star, implying different optical depth for
different Balmer lines. In that scenario the H$_{\alpha}$ line is
formed  further out from the star compared to H$_{\beta}$ or
H$_{\eta}$ (see Fig. 15 in \citep{2001ApJ...553..837H}).
As the event develops, due either to wind--wind collision in a
binary system or to a shell ejection of the primary star,  the
optical depth in the line of sight increase and  P~Cygni absorption
can form in the H$_{\alpha}$ line. Viewed, however, from the poles
at FOS4, the optical depth is always high enough to show a P~Cygni
profile.

\acknowledgments{}
KW is supported  by  the state of Northrhine-Westphalia 
(Lise-Meitner fellowship). DJB acknowledges support by the 
DFG SFB\ 591 `Universal Processes in Non-equilibrium Plasmas'. 
We thank Wolfgang J. Duschl for discussion and comments on the paper. 
We are grateful to the referee for valueable remarks and suggestions.

\clearpage

\begin{figure}[t]
\includegraphics[width=8cm]{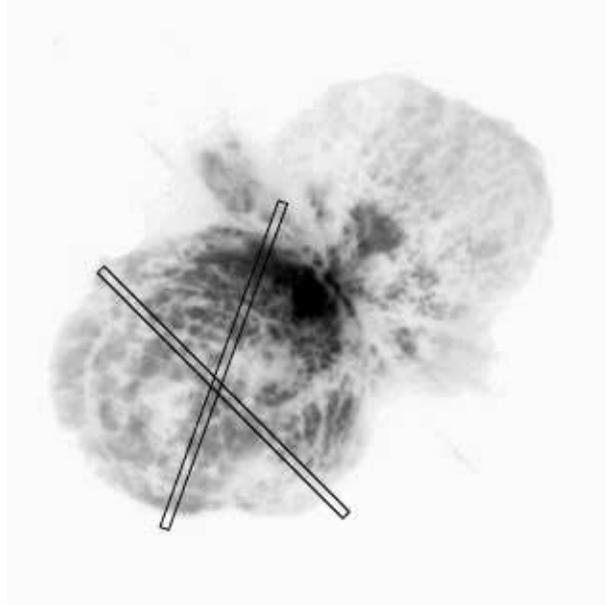}
\hspace{8cm}

\caption{HST ACS/HRC image in the F550M filter of $\eta$ Carinae and
the Homunculus. North  is up and east to the left, both slit
positions (PA=160$\degr$ and  PA=45$\degr$) are indicated. The slits
are drawn to their full (maximum possible) width and length. FOS4 is
an area of $\approx$ 1\arcsec\,$\times$\,1\arcsec\ located
approximately at the intersection of the two slits. }
\label{fig_fos4}
\end{figure}

\begin{figure}[t]
\includegraphics[angle=270,width=8cm]{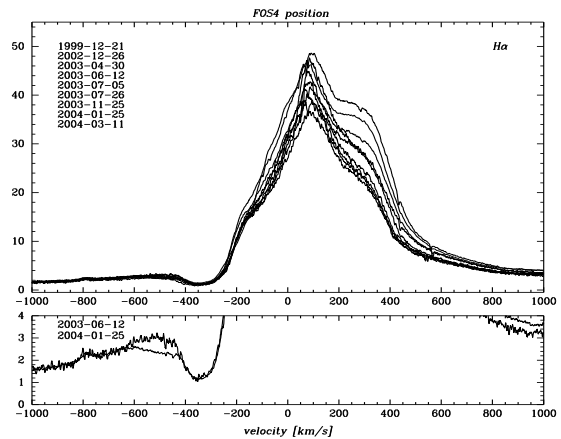}
\caption{The  H$_{\alpha}$
profile from some of the spectra.
The lower panel shows the changes in the P~Cygni absorption
for two of the spectra.
}
\label{fig_ha}
\end{figure}

\begin{figure}[t]
\includegraphics[angle=270,width=8cm]{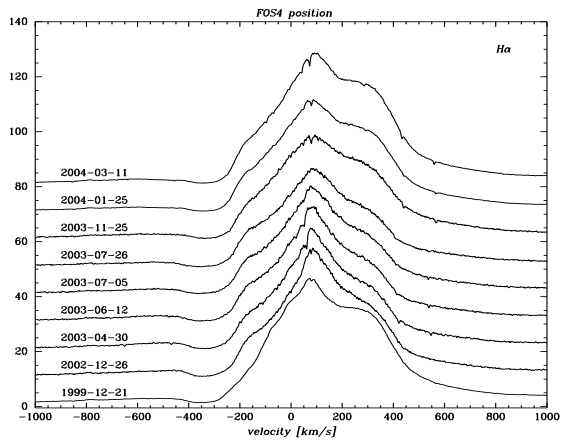}
\caption{ H$_{\alpha}$ profiles of selected spectra at position
FOS4. In order to avoid confusion, each spectrum is offset by
additional 10 continuum units. Spectra are labeled with the
observing date. None of the spectra shows a clear flat-topped
profile. Note also the evolution of the red hump with time.}
\label{fig_shift}
\end{figure}

\begin{figure}[t]
\includegraphics[angle=270,width=8cm]{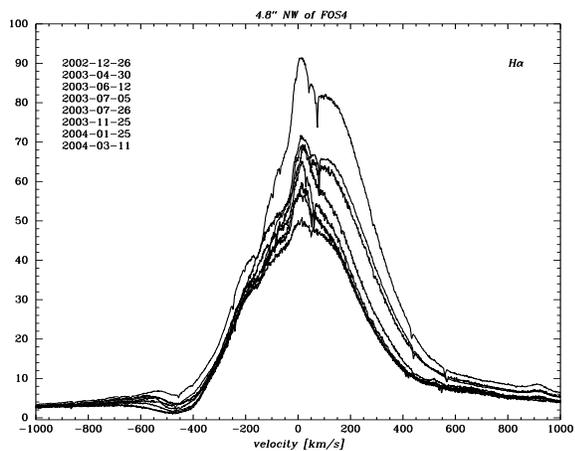}
\hspace{8cm}
\caption{H$_{\alpha}$
profile from the same  spectra  as in Fig. \ref{fig_ha}
but now at a  position 4\farcs8 north-west of FOS4, closer
to the central source. Since the 1999 data had a different
position  angle, this spectrum  is missing here.
} \label{fig_ha48}
\end{figure}

\begin{figure}[t]
\includegraphics[angle=270,width=8cm]{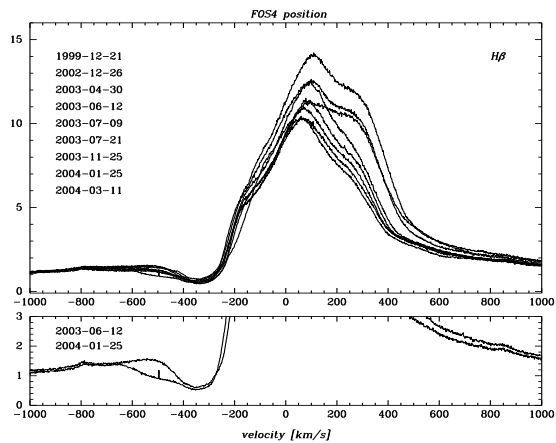}
\caption{Same as Fig.\ \ref{fig_ha} but for H$_{\beta}$.}
\label{fig_hb}
\end{figure}

\begin{figure}[t]
\includegraphics[angle=270,width=8cm]{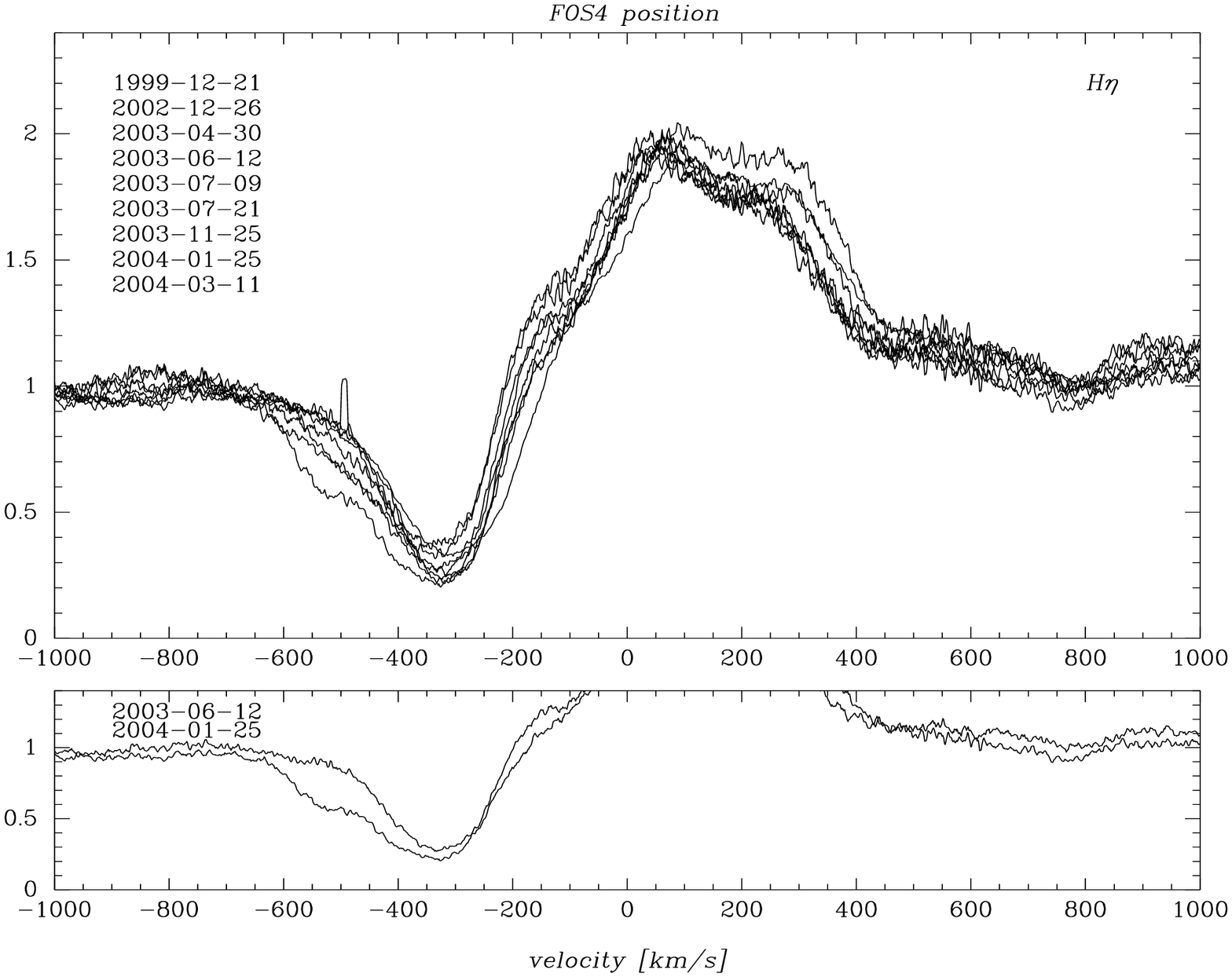}
\caption{Same as Fig.\ \ref{fig_ha} but for H$_{\eta}$.}
\label{fig_hh}
\end{figure}

\begin{figure}[t]
\includegraphics[angle=270,width=8cm]{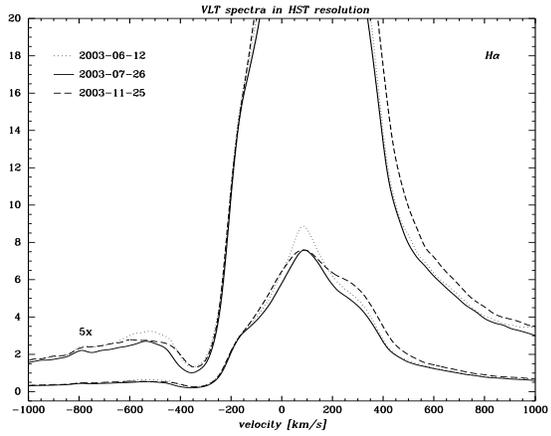}
\caption{H$_{\alpha}$ profile observed at FOS4 with the VLT, but
convolved to the spectral resolution of the HST/STIS data. The plot
and dates are chosen to be directly comparable to Fig. 1c of the
accompanying paper of Davidson et al.} \label{fig_hstvlt}
\end{figure}

\clearpage
\renewcommand{\arraystretch}{.6}

\begin{table}
\caption{Observing dates. Phase has been computed with zero point
of MJD = 50800 and a period of 2023 days.}
\label{tab_obs}
\begin{center}
\begin{tabular}{|l|l|l|l|}
\tableline
\multicolumn{2}{|c|}{Date} & MJD & phase \\
\tableline
1999/12/21 & 1999.961 & 51534 & 0.363 \\
2002/12/08 & 2002.937 & 52617 & 0.898 \\
2002/12/26 & 2002.986 & 52635 & 0.907 \\
2002/12/31 & 2003.000 & 52640 & 0.910 \\
2003/01/03 & 2003.008 & 52643 & 0.911 \\
2003/01/19 & 2003.052 & 52659 & 0.919 \\
2003/01/23 & 2003.063 & 52664 & 0.921 \\
2003/02/04 & 2003.100 & 52675 & 0.927 \\
2003/02/14 & 2003.123 & 52685 & 0.932 \\
2003/02/25 & 2003.153 & 52696 & 0.937  \\
2003/03/07 & 2003.180 & 52706 & 0.942 \\
2003/03/12 & 2003.195 & 52711 & 0.945 \\
2003/04/30 & 2003.329 & 52760 & 0.969 \\
2003/05/05 & 2003.342 & 52765 & 0.971 \\
2003/05/12 & 2003.361 & 52772 & 0.975 \\
2003/05/29 & 2003.408 & 52789 & 0.983 \\
2003/06/03 & 2003.423 & 52794 & 0.986 \\
2003/06/08 & 2003.436 & 52799 & 0.988 \\
2003/06/12 & 2003.447 & 52803 & 0.990 \\
2003/06/17 & 2003.460 & 52808 & 0.993 \\
2003/06/22 & 2003.474 & 52812 & 0.995 \\
2003/06/30 & 2003.496 & 52820 & 0.999 \\
2003/07/05 & 2003.510 & 52825 & 1.001 \\
2003/07/09 & 2003.521 & 52829 & 1.003 \\
2003/07/16 & 2003.540 & 52836 & 1.007 \\
2003/07/20 & 2003.551 & 52840 & 1.009 \\
2003/07/26 & 2003.567 & 52846 & 1.012 \\
2003/07/31 & 2003.581 & 52851 & 1.014 \\
2003/11/25 & 2003.901 & 52968 & 1.072 \\
2003/12/17 & 2003.962 & 52991 & 1.083 \\
2004/01/02 & 2004.005 & 53007 & 1.091 \\
2004/01/25 & 2004.068 & 53030 & 1.102 \\
2004/02/20 & 2004.140 & 53056 & 1.115 \\
2004/03/11 & 2004.192 & 53076 & 1.125 \\
\tableline
\end{tabular}
\end{center}
\end{table}

\end{document}